\pdfoutput=1
\documentclass[aps,prl,twocolumn,showpacs,superscriptaddress]{revtex4-1}
\usepackage{graphicx}

\newcommand{\Nbar}{\overline{N}}
 
\renewcommand{\vec}[1]{\mathbf{#1}}

\newcommand{\vs}{{\it vs.\ }}

\newcommand{\ie}{{\it i.~e.,\ }}

\begin{document}
\title{Universality of Block Copolymer Melts}



\author{Jens~Glaser}
\altaffiliation{current affiliation: Department of Chemical Engineering, University of Michigan, Ann Arbor, MI 48109, USA}
\author{Pavani~Medapuram}
\affiliation{Department of Chemical Engineering and Materials Science, University of Minnesota, 421 Washington Ave SE, Minneapolis, MN 55455, USA}
\author{Thomas~M.~Beardsley}
\author{Mark~W.~Matsen}
\altaffiliation{current affiliation: Institute for Nanotechnology, University of Waterloo, QNC 5602, Waterloo, Ontario, N2L 3G1, Canada}
\affiliation{School of Mathematical and Physical Sciences, University of Reading, Whiteknights, Reading RG6 6AX, U.K.}
\author{David C. Morse}
\email[Corresponding author, email:\ ] {morse012@umn.edu}
\affiliation{Department of Chemical Engineering and Materials Science, University of Minnesota, 421 Washington Ave SE, Minneapolis, MN 55455, USA}

\date{\today}

\begin{abstract}
Simulations of five different coarse-grained models of symmetric diblock copolymer melts are compared to demonstrate a universal (\ie model-independent) dependence of the free energy on the invariant degree of polymerization $\Nbar$, and to study universal properties of the order-disorder transition (ODT). The ODT appears to exhibit two regimes: Systems of very long chains ($\Nbar \agt 10^{4}$) are well described by the Fredrickson-Helfand theory, which assumes weak segregation near the ODT. Systems of smaller but experimentally relevant values, $\Nbar \alt 10^4$, undergo a transition between strongly segregated disordered and lamellar phases that, though universal, is not adequately described by any existing theory.
\end{abstract}


\maketitle

Universality is a powerful feature of polymer statistical mechanics that allows the behavior of real systems to be predicted on the basis of simple generic models and scaling arguments.  The paradigmatic example is the scaling theory of dilute and semidilute polymer solutions in good solvents \cite{deGennes_book_79, Schaefer_99, Graessley_04}, which predicts a universal dependence of all properties on two thermodynamic state parameters (an excluded volume parameter and an overlap parameter). Historically, this scaling hypothesis was verified by comparing experiments on diverse chemical systems with varied chain lengths and concentrations \cite{Daoud_deGennes_75,Noda_81,Graessley_04}. Here, we compare simulations of diverse models to verify an analogous scaling hypothesis about the equation of state and order-disorder transition (ODT) of symmetric diblock copolymers, and to characterize this transition.

We consider a dense liquid of AB diblock copolymers, with N monomers per chain, and a fraction $f_{A}$ of A monomers. We focus on the symmetric case, $f_{A}=1/2$. Self-consistent field theory (SCFT) is the dominant theoretical approach for block copolymers \cite{Leibler_80, Matsen_Schick_94a, Matsen_02}. SCFT describes polymers as random walks with a monomer statistical segment length $b$, which we take to be equal for A and B monomers. The free energy cost of contact between A and B monomers is characterized by an effective Flory-Huggins interaction parameter $\chi_{e}$. Let $g$ denote a dimensionless excess free energy per chain, normalized by the thermal energy $k_{B}T$. SCFT predicts a free energy $g$ for each phase that depends only upon $f_{A}$ and the product $\chi_e N$, or upon $\chi_e N$ alone for $f_A=1/2$. This yields a predicted phase diagram \cite{Leibler_80, Matsen_Schick_94a} that likewise depends only on $f_{A}$ and $\chi_e N$. For $f_{A}=1/2$, SCFT predicts a transition between the disordered phase and lamellar phase at $(\chi_e N)_{\mathrm{ODT}} = 10.495$. 

SCFT is believed to be exact in the limit of infinitely long, strongly interpenetrating polymers \cite{deGennes_77, Fredrickson_Helfand_87}. The degree of interpenetration in a polymer liquid is characterized by a dimensionless concentration $\overline{C} \equiv cR^{3}/N$, in which $c$ is monomer concentration, $c/N$ is molecule concentration, and $R = \sqrt{N}b$ is coil size. Alternatively, interpenetration may be characterized by the invariant degree of polymerization $\Nbar \equiv \overline{C}^{2} = N(cb^{3})^2$ \cite{Fredrickson_Helfand_87}. A series of post-SCF theories \cite{Fredrickson_Helfand_87, Barrat_Fredrickson_91, deLaCruz_91, Mayes_deLaCruz_91, Holyst_Vilgis_93, Beckrich_07, Grzywacz_Morse_07, Qin_Morse_11, Morse_Qin_11}, starting with the Fredrickson-Helfand (FH) theory \cite{Fredrickson_Helfand_87}, have given predictions for {\it finite} diblock copolymers that depend on $\Nbar$ in addition to the SCFT state parameters, but that reduce to SCFT predictions in the limit $\Nbar\rightarrow \infty$. Specifically, these theories suggest that, for symmetric copolymers, $g$ of each phase is given by a universal (model- and chemistry-independent) function of $\chi_e N$ and $\Nbar$ alone,
\begin{equation}
   g = g(\chi_e N, \Nbar)
   \quad. \label{eq:g_scale}
\end{equation}
If so, the value of $\chi_e N$ at the ODT (where the free energies of the two phases are equal) should depend on $\Nbar$ alone, and should approach 10.495 as $\Nbar \rightarrow \infty$. 

The FH theory and its relatives all yield predictions that are consistent with Eq.~(\ref{eq:g_scale}). All, however, also rely on mathematical approximations that, to a greater or lesser extent \cite{Morse_Qin_11}, limit their validity to large values of $\Nbar$. Validity of Eq.~(\ref{eq:g_scale}) is thus a necessary but not sufficient condition for validity of more approximate theories. One goal of this work is to directly test the validity of this scaling hypothesis, independent of the FH theory, by comparing results of simulations of different coarse-grained models.

Here, we compare simulations of four different continuum bead-spring models (models H, S1, S2, and S3) and a lattice model (model F) with widely varying chain lengths. Each bead-spring model has a pair potential of the form $V_{ij}(r) = \epsilon_{ij}u(r)$, with $\epsilon_{AA}=\epsilon_{BB}$ and $\epsilon_{AB} \geq \epsilon_{AA}$. Model H  uses a truncated purely repulsive Lennard-Jones pair potential (H denotes ``hard"), and is similar to the model of Grest and coworkers \cite{Grest_Kremer_96, Grest_Kremer_99}. Models S1, S2, and S3 all use the softer pair potential typical of dissipative particle dynamics simulations. Model F is an FCC lattice model. Models H \cite{Qin_Morse_12, Glaser_Morse_12, Glaser_Morse_14}, S1 \cite{Glaser_Morse_12, Glaser_Morse_14}, and F \cite{Vassiliev_Matsen_03, Matsen_Vassiliev_06, Beardsley_Matsen_10, Beardsley_Matsen_11} have been studied previously. The term ``model" refers to set of choices for the functional form of the pair and bond potentials, and for values of all parameters except $N$ and one parameter that is varied to control $\chi_e$. Here, we vary the difference $\alpha \equiv \epsilon_{AB} - \epsilon_{AA}$ between the strength of $AB$ and $AA$ (or $BB$) pair interactions, while holding $T$, $\epsilon_{AA}$ and other parameters constant.

The parameters of the four bead spring models were chosen to facilitate testing of universality, by creating pairs of simulations of different models with equal values of $\Nbar$. Parameters for models H, S1, S2, and S3 were adjusted to give values of $\Nbar/N =(cb^{3})^{2}$ with ratios of nearly 1:4:16:32. Because simulations were conducted for chain lengths $N=$ 16, 32, 64, and 128 that also differ by multiples of 2, some pairs of simulations of different models have nearly equal values of $\Nbar$. Specifically, simulations of H-64 (model H with $N$=64) and S1-16 (model S1 with $N$=16) both have $\Nbar \simeq 240$, while S1-64 and S2-16 both have $\Nbar \simeq 960$, S1-128, S2-32, and S3-16 all have $\Nbar \simeq 1920$, and S3-64 and S2-32 both have $\Nbar \simeq 3840$.

The simulations presented here span a range $\Nbar \simeq 100 - 7600$ that overlaps much of the range of $\Nbar \simeq 200 - 20,000$ explored in experiments on symmetric diblock copolymers. For example: $\Nbar \simeq$ 1100 in a classic study of poly(styrene-{\it b}-isoprene) \cite{Bates_Mortensen_94,Khandpur_Bates_95}, $\Nbar \simeq$ 220 in a recent study of poly(isoprene-{\it b}-L lactic acid) \cite{Lee_Bates_13}, and $\Nbar \simeq$ 5000 in the study of poly(ethylene-propylene-{\it b}-ethylethylene) used to test the FH theory \cite{Bates_Rosedale_88,Bates_Rosedale_90,Rosedale_Bates_95,Qin_thesis_09}.

{\it Simulation details}: Each bead-spring model has a bond potential $V_{\mathrm{bond}}(r) = \kappa( r - l_{0})^2/2$ and a pair potential $V_{ij}(r) = \epsilon_{ij}u(r)$ that vanishes beyond a cutoff distance $r_{c}$. For model H, $u(r) = 4[ (\sigma/r)^{-12} - (\sigma/r)^{-6} + 1/4]$, $r_{c} = 2^{1/6}\sigma$, $\epsilon_{AA}=k_{B}T$, $\kappa = 400 k_{B}T/\sigma^2$ and $l_{0} = \sigma$. For models S1, S2, and S3, $u(r) = [1 - (r/\sigma)^{2}]/2$, $r_{c} = \sigma$, $\epsilon_{AA}=25 k_{B}T$, $l_{0}=0$, and $\kappa \sigma^{2}/k_B T =$ 3.406, 1.135, and 0.867 respectively. All bead-spring simulations reported here are GPU-accelerated NPT molecular dynamics (MD) simulations \cite{Anderson2008}. The pressure for each such model (which is independent of $N$) was chosen to yield a target monomer concentration $c$ in the limit $\alpha = 0$, $N \rightarrow \infty$ of infinite homopolymers. Values of $c\sigma^{3}$ for models H, S1, S2, and S3 are 0.7, 3.0, 1.5, and 1.5, respectively. Values of the statistical segment length $b$ for all five models were obtained \cite{Morse_Chung_09, Glaser_Morse_14} from the extrapolation $b^{2} \equiv \lim_{N \rightarrow \infty} 6R_{g}^{2}/N$ of homopolymer ($\alpha = 0$) simulations, where $R_{g}$ is the radius of gyration, giving $b/\sigma =$ 1.404, 1.088, 1.727, 1.938 for models H, S1, S2, and S3, respectively. Model F is an FCC lattice model with 20~\% vacancies, with a bond length $\sqrt{2}d$, $c = 0.4 d^{-3}$, and $b = 1.745 d$. ODTs for bead-spring models were identified using a well-tempered metadynamics free energy method, as discussed in supplemental material \cite{supplementary}.

{\it Estimating $\chi_{e}$}: The question of how to assign a value to the interaction parameter $\chi_e$ used in coarse-grained theories, and in Eq (\ref{eq:g_scale}), has hindered previous attempts to compare simulations of coarse-grained models to theory or to each other. For each model in our simulations, $\chi_e$ is some unknown function $\chi_e(\alpha)$ of the control parameter $\alpha$. Our approach to estimating $\chi_{e}(\alpha)$ is motivated by recent advances theoretical predictions for the structure factor $S(q)$ in the disordered phase. It was recently shown \cite{Glaser_Morse_14} that the renormalized one-loop (ROL) theory \cite{Grzywacz_Morse_07, Qin_Morse_11} can accurately predict simulation results for $S(q)$ over a wide range of values of $N$ and $\alpha$ using a single nonlinear function $\chi_{e}(\alpha)$ for all chain lengths. In what follows, we thus analyze results for the free energy and ODT using an approximation for $\chi_{e}(\alpha)$ for each model that is obtained from a simultaneous fit of simulation results for $S(q)$ for several chain lengths to the ROL theory (see supplemental material \cite{supplementary}).

For comparison, we also consider a simpler linear approximation for $\chi_e(\alpha)$, which was used in Ref.~\cite{Qin_Morse_12}. This approximation is based on an analysis of a perturbation theory for homopolymer blends \cite{Morse_Chung_09} that yields an exact expression for the first term of a Taylor expansion of $\chi_{e}(\alpha)$. This gives $\chi_{e}(\alpha) \simeq z_{\infty}\alpha/k_{B}T$, where the coefficient $z_{\infty}$ is obtained from homopolymer ($\alpha=0$) simulations \cite{supplementary}.

{\it Results}: One rather direct way of testing Eq.~(\ref{eq:g_scale}) is to compare results from different simulation models for the derivative $g' \equiv \partial g / \partial (\chi_e N)$. Given an accurate estimate of $\chi_{e}(\alpha)$, $g'$ can be calculated using the relation
\begin{equation}
    \frac{\partial g}{\partial (\chi_e N)}
    = \frac{\langle U_{AB}(\alpha) \rangle}{ MN \epsilon_{AB}(\alpha)}
      \left [ k_B T \frac{d\chi_e(\alpha)}{d\alpha} \right ]^{-1}
    , \label{eq:dg_dchiN}
\end{equation}
where $U_{AB}$ is the total nonbonded AB pair interaction energy in a system of $M$ chains. Eq.~(\ref{eq:dg_dchiN}) is derived by using the identity $\partial g/\partial \alpha = \langle \partial H(\alpha)/\partial \alpha \rangle/ (k_B T M)$, where $H(\alpha)$ is the model Hamiltonian, to show that $\partial g/\partial \alpha = \langle U_{AB} \rangle/(M k_B T \epsilon_{AB})$, and then writing $\partial g/\partial\chi_e = (\partial g/\partial \alpha) / (d\chi_e/d\alpha)$.

\begin{figure}
\includegraphics[width=\columnwidth]{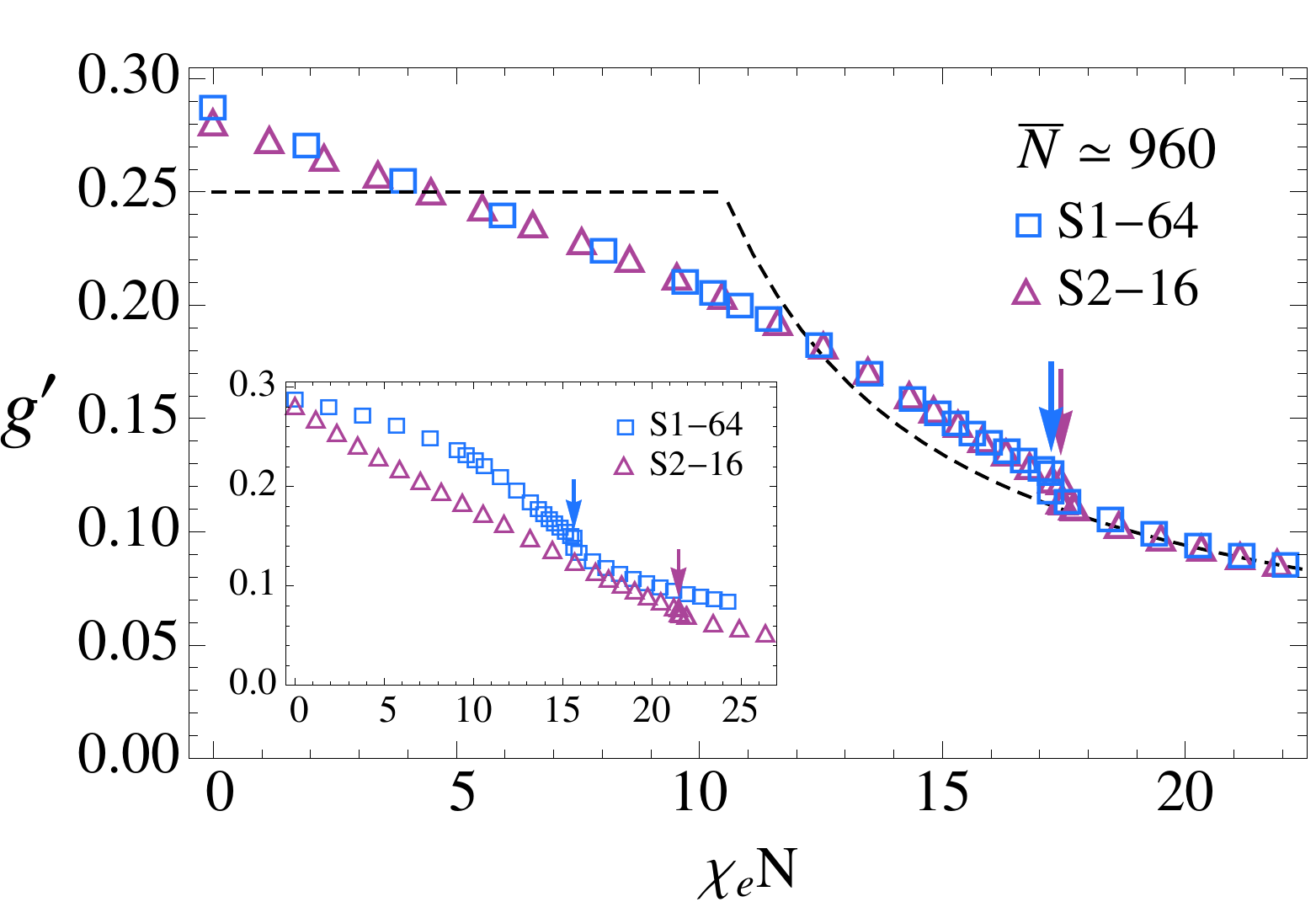}
\caption{Plots of $g' \equiv \partial g/\partial (\chi_e N)$ vs. $\chi_e N$ for models S1-64 and S2-16 ($\Nbar \simeq 960$), constructed using different approximations for $\chi_{e}(\alpha)$: The inset uses the linear approximation $\chi_{e}(\alpha) = z_{\infty}\alpha/k_B T$. The main plot uses a nonlinear approximation obtained by fitting $S(q)$. The dashed curve shows the SCFT prediction for $g'(\chi_e N)$. Vertical arrows mark the positions of the ODTs. \label{fig:lin-nonlin-theta}}
\end{figure}

Eq.~(\ref{eq:g_scale}) implies that $g' = \partial g/\partial (\chi_e N)$ should (like $g$) be a universal function of $\chi_e N$ and $\Nbar$. Data from simulations of different models with matched values of $\Nbar$ should thus collapse when $g'$ is plotted vs. $\chi_{e}N$. The quality of the collapse does, however, depend on the accuracy of the approximation for $\chi_{e}(\alpha)$ used to construct such a plot. The inset and main plots of Fig.~\ref{fig:lin-nonlin-theta} show two different attempts to collapse data for $g'$ vs. $\chi_e N$ for models S1-64 and S2-16, for which $\Nbar \simeq 960$. The inset was constructed using the linear approximation $\chi_{e} \simeq z_{\infty}\alpha/k_{B}T$. This approach fails, yielding a poor data collapse and poor agreement for the value $(\chi_e N)_{\mathrm{ODT}}$ of $\chi_e N$ at the ODT (indicated by arrows). The main plot was constructed using the nonlinear approximation for $\chi_{e}(\alpha)$ obtained by fitting $S(q)$. This succeeds, giving near perfect collapse of the data for $g'$ from these two models, and excellent agreement for $(\chi_{e}N)_{\mathrm{ODT}}$. Results for other pairs of models with matching $\Nbar$ show similar agreement. This analysis nicely verifies the accuracy of both the scaling hypothesis, Eq.~(\ref{eq:g_scale}), and of this method of estimating $\chi_e(\alpha)$.

There is a small discontinuity in $g'$ across the ODT in the main plot of Fig.~\ref{fig:lin-nonlin-theta}, of magnitude $\Delta g' \simeq 0.007$, indicating a very weakly first-order transition. The smallness of $\Delta g'$ indicates that the degree of AB contact is similar in the disordered and ordered phases near the ODT. This suggests that the disordered phase at the ODT has a local structure similar to that of the ordered phase, with well defined A and B domains and an AB interfacial area per volume very similar to that of the lamellar phase, but without long range order. The SCFT prediction for $g'(\chi_e N)$ (dashed line) is given by the spatial average of the product $\phi_{A}({\vec r})\phi_{B}({\vec r})$ of the predicted local volume fractions of A and B monomers. This yields $g' =$ 0.25 in the disordered phase, $\chi_e N < 10.495$. Notably, SCFT predictions for $g'$ are poor in the disordered phase, but show excellent agreement with simulations in the ordered phase. SCFT thus accurately predicts the extent of $AB$ contact within the ordered phase, but is intrinsically incapable of handling the strong short-range correlations in the disordered phase.

\begin{figure}[t]
\includegraphics[width=\columnwidth]{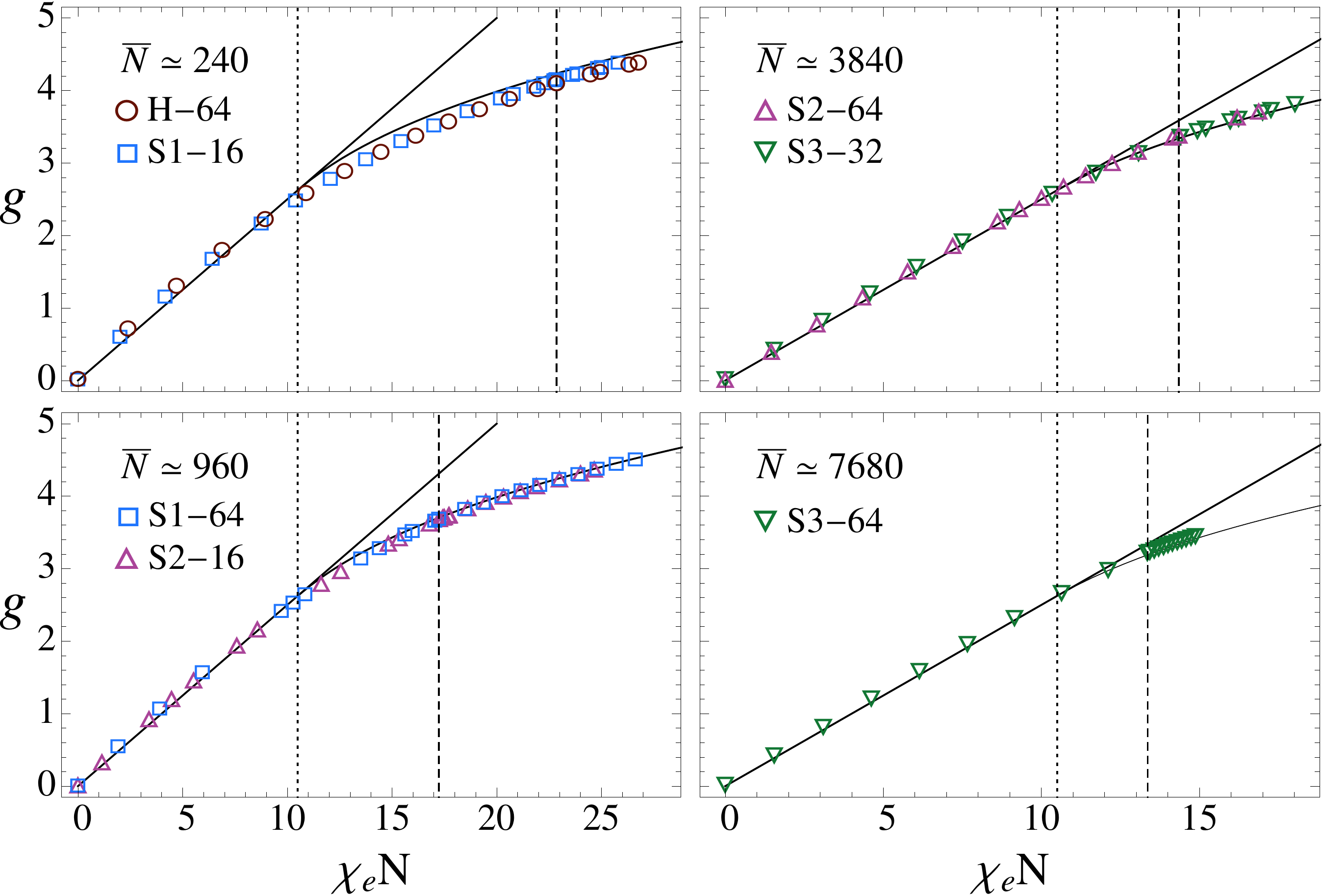}
\caption{Free energy per chain $g$ vs. $\chi_e N$ at 4 different values of $\Nbar$, plotted using a nonlinear approximation for $\chi_{e}(\alpha)$. Solid lines are SCFT predictions for $g(\chi_e N)$. The straight solid line is the SCFT prediction $g(\chi_e N) = \chi_e N/4$ for the disordered phase. Vertical dotted lines show the SCFT ODT, at $\chi_e N = 10.495$. Vertical dashed lines show actual ODTs. In plots that display results for two systems, the ODT is shown for the system with larger $N$. \label{fig:g_vs_chiN}}
\end{figure}

Fig.~\ref{fig:g_vs_chiN} shows the free energy per chain $g$ \vs $\chi_{e} N$ for four values of $\Nbar$. These were calculated by numerically integrating simulation results for $\partial g/\partial \alpha$ within each phase, setting $g(\alpha=0)=0$ by convention for homopolymers, and matching values of $g$ in the two phases at the ODT. Three of the plots show results for pairs of simulations with matched values of $\Nbar$, to demonstrate consistency of results obtained in corresponding thermodynamic states of different models. Deviations from the SCFT prediction for $g$ in the disordered phase are easily visible in the range $10.495 < \chi_e N < (\chi_e N)_{\mathrm{ODT}}$ between the SCFT and true ODTs, where the disordered phase develops strong correlations. Interestingly, SCFT predictions for $g$ are quite accurate within the ordered phase, and become more so with increasing $\Nbar$:  There is a small but noticeable offset between simulation results and SCFT predictions for $g$ in the ordered phase for $\Nbar \simeq 240$, but much less error for larger $\Nbar$. This agreement does not follow trivially from the observed accuracy of SCFT predictions for $g'$ in the ordered phase, since the value of $g$ at the ODT is calculated by integrating $\partial g/\partial \alpha$ through the disordered phase, in which SCFT predictions are poor. At a heuristic level, the main components of $g$ are free energies arising from AB interfacial contact and chain stretching. Only the extent of AB contact is directly reflected by the value of $g'$. These results thus suggest that SCFT accurately describes both of these free energy components in the ordered phase, though not in the disordered phase near the ODT.

\begin{figure}[t]
\centering\includegraphics[width=\columnwidth]{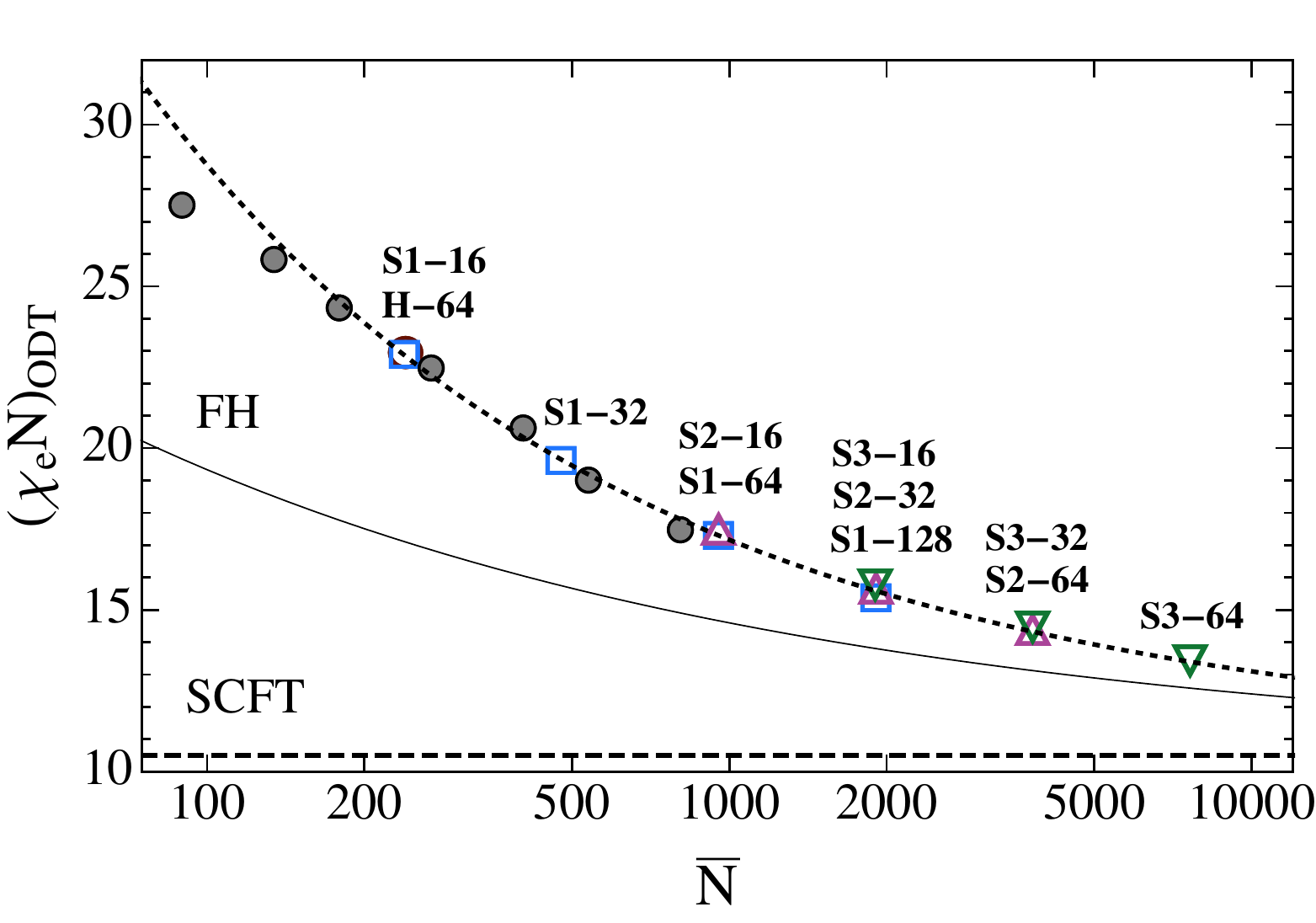}
\caption{Values of $\chi_{e}N$ at the ODT \vs $\Nbar$, for all simulations. Bead-spring model results are shown as open symbols, with labels for specific systems. Lattice model results for $N=$ 20, 30, 40, 60, 90, 120 and 180 are filled gray circles. The solid curve is the FH prediction. The horizontal long dashed line is the SCFT prediction. The short dashed curve is an empirical fit. \label{fig:odt_summary}}
\end{figure}

Fig.~\ref{fig:odt_summary} shows a compilation of results for $(\chi_e N)_{ODT}$ from all simulations, plotted vs. $\Nbar$, using our nonlinear approximation for $\chi_{e}(\alpha)$. The most important feature of this plot is the fact that results from all five models collapse onto a common curve, as required by Eq.~(\ref{eq:g_scale}), confirming the universality of the results. Note the excellent agreement found for pairs of simulations with matched values of $\Nbar$, shown by overlapping open symbols. The results also clarify the limitations of the FH prediction \cite{Fredrickson_Helfand_87}, $(\chi_e N)_{\mathrm{FH}} \equiv 10.495 + 41.0 \Nbar^{-1/3}$ (solid curve). The highest values of $\Nbar$ studied here closely approach the FH prediction, but deviations grow with decreasing $\Nbar$, and become large for modest values typical of many experiments. The dotted curve is an empirical fit to results of the bead spring models: $(\chi_e N)_{\mathrm{ODT}} = (\chi_e N)_{\mathrm{FH}} + 123.0 \Nbar^{-0.56}$. These results suggest that the FH theory becomes accurate for $\Nbar \agt 10^{4}$, but breaks down at lower $\Nbar$.

\begin{figure}[t]
\centering\includegraphics[width=\columnwidth]{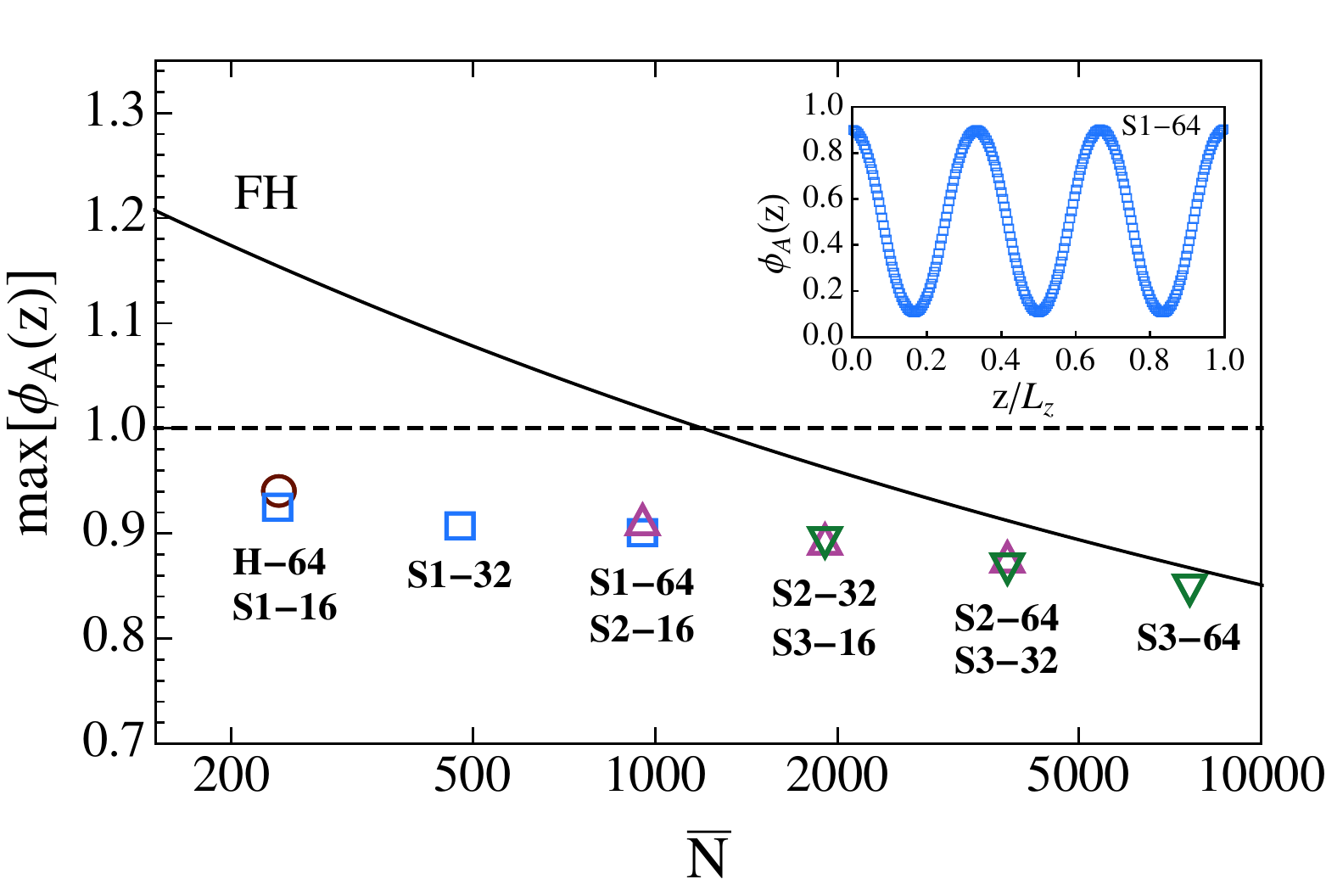}
\caption{Maximum value of the local volume fraction $\phi_{A}(z)$ in the ordered phase at the ODT plotted \vs $\Nbar$ for all bead spring simulations. Inset: Composition profile $\phi_{A}(z)$ in the ordered phase at the ODT for model S1-64, where $z$ is distance normal to layers and $L_{z}$ is simulation cell size. \label{fig:amplitude_vs_nbar}}
\end{figure}

Insight into the reason for this breakdown of the FH theory for $\Nbar \alt 10^4$ can be gained by examining the degree of segregation in the ordered phase at the ODT. The approximations underlying the FH theory are strictly valid only for extremely large $\Nbar$, for which it predicts a transition to a weakly segregated lamellar phase. The inset of Fig.~\ref{fig:amplitude_vs_nbar} shows the dependence of the average local volume fraction of A monomers, $\phi_{A}(z)$, in the ordered phase at the ODT, plotted \vs normal coordinate $z$ for model S1-64 ($\Nbar \simeq 960$). This composition profile is almost sinusoidal, but is clearly not weakly segregated. The main plot shows the maximum value of $\phi_{A}(z)$ in the middle of the A domain of the ordered phase at the ODT plotted vs. $\Nbar$ for different systems. This value remains large ($\geq 0.83$) over the entire range studied here, but decreases slowly with $\Nbar$ in a manner that suggests convergence to FH predictions for $\Nbar \agt 10^{4}$. The solid curve shows the corresponding value predicted by the FH theory, which assumes a sinusoidal profile. Note that the FH theory predicts unphysical values of $\max[\phi_{A}(z)] > 1$ for $\Nbar \alt 10^{3}$, and thus {\it must} begin to fail below a crossover value of $\Nbar$ somewhat greater than $10^{3}$.

In this Letter, we present the first simulations to demonstrate consistency among different coarse-grained models for the equation of state and the value of $\chi_e N$ at the ODT of symmetric diblock copolymers, by using $\Nbar$ as a correlating variable. This verifies a scaling hypothesis, Eq.~(\ref{eq:g_scale}), that posits a universal dependence on $\Nbar$ and $\chi_e N$. At a practical level, the demonstration of universality opens the way for the use of coarse-grained simulations as reliable tools for predicting the behavior of real materials. This success depended critically upon the development of an adequate method of estimating $\chi_{e}$, which we achieve by fitting the structure factor $S(q)$ in the disordered phase to an accurate new theory \cite{Grzywacz_Morse_07,Qin_Morse_11}. The universality predicted by Eq.~(\ref{eq:g_scale}) is found to be remarkably robust, applying down to $\Nbar \simeq 200$, and to chains with as few as 16 monomers. Universal behavior characteristic of random-walk polymers must, of course, break down for sufficiently short discrete chains, but we found surprisingly little evidence of this in the systems studied here. The FH theory has a more limited range of validity. Our results suggests that the FH theory becomes quantitatively accurate for $\Nbar \agt 10^{4}$. In the range $\Nbar \alt 10^4$ studied here (and in many experiments), both the ordered phase and the disordered phase become strongly segregated near the ODT. This violates the assumptions underlying the FH theory, causing $(\chi_e N)_{\mathrm{ODT}}$ to deviate substantially from the FH prediction. SCFT is found, however, to give surprisingly accurate predictions for $g$ in the ordered phase. SCFT thus may provide good predictions for many order-order transitions in block copolymers, if combined with sufficiently accurate estimates of $\chi_e$.  On the other hand, SCFT grossly underestimates $(\chi_e N)_{\mathrm{ODT}}$ for the order-disorder transition of symmetric diblock copolymers with modest $\Nbar$ because it cannot describe the strongly segregated disordered phase that exists near the ODT.

\begin{acknowledgments}
We acknowledge fruitful discussions with Juan de Pablo and Jian Qin, and thank Michael Engel for carefully reading the manuscript. This work was supported by NSF awards DMR-0907338 and DMR-1310436, EPSRC (EP/E10342/1), a DFG postdoctoral fellowship for J.G. (GL 733/1-1), and a Univ. of Minnesota doctoral fellowship for P.M. The research used resources of the Minnesota Supercomputing Institute and of the Keeneland Computing Facility, which is supported by NSF Contract OCI-0910735.
\end{acknowledgments}

\end{document}

\title{Universality of Block Copolymer Melts: Supplemental Material}

\author{Jens~Glaser}
%\altaffiliation{current affiliation: Department of Chemical Engineering, University of Michigan, Ann Arbor, MI 48109, USA}
\author{Pavani~Medapuram}
\affiliation{Department of Chemical Engineering and Materials Science, University of Minnesota, 421 Washington Ave SE, Minneapolis, MN 55455, USA}
\author{Thomas~M.~Beardsley}
\author{Mark~W.~Matsen}
%\altaffiliation{current affiliation: Institute for Nanotechnology, University of Waterloo, QNC 5602, Waterloo, Ontario, N2L 3G1, Canada}
\affiliation{School of Mathematical and Physical Sciences, University of Reading, Whiteknights, Reading RG6 6AX, U.K.}
\author{David~C.~Morse}
%\email[Corresponding author, email:\ ] {morse012@umn.edu}
\affiliation{Department of Chemical Engineering and Materials Science, University of Minnesota, 421 Washington Ave SE, Minneapolis, MN 55455, USA}

%\date{\today}

%\begin{abstract}
%Supplementary Material
%\end{abstract}

%\pacs{}% insert suggested PACS numbers in braces on next line

\maketitle

\section{Introduction}
The simulations presented in the accompanying article were conducted independently by two subsets of the authors: Simulations of the bead-spring models (models H, S1, S2, and S3) were conducted at the University of Minnesota by J.G., P.M. and D.M, while the lattice Monte Carlo simulations (model F) were conducted at the University of Reading by T.B. and M.M. The decision to present the results in a joint publication was made when analysis of the two data sets by identical methods revealed the consistency of the results, after both sets of simulations were complete. Somewhat different simulation methods were thus used in the bead-spring and lattice simulations. 

\section{FCC Lattice Model}

Model F is a face-centered-cubic lattice Monte Carlo model with 20 \% vacancies, as described in the main text. The model has a distance $\sqrt{2}d$ between nearest neighbor sites and a concentration $c = 0.4 d^{-3}$, where $d$ is the spacing of an underlying cubic lattice. Double occupancy is prohibited. The model has no interaction between $AA$ or $BB$ nearest neighbor pairs ($\epsilon_{AA} = \epsilon_{BB} = 0$), and an interaction $\epsilon_{AB} = \alpha$ between AB nearest neighbor pairs. 

This model has been studied previously \cite{Vassiliev_Matsen_03, Matsen_Vassiliev_06, Beardsley_Matsen_10, Beardsley_Matsen_11} using methods similar to those used here. The simulations reported here all used the parallel tempering method described in Ref.~\cite{Beardsley_Matsen_10}. Simulations were carried out in $L \times L \times L$ cubic simulation cells, where $L$ was chosen to be approximately $\sqrt{14}$ times the preferred layer spacing at the ODT, to yield a nearly commensurate cell for lamellae oriented in a $\{321\}$ direction. Previous studies \cite{Matsen_Vassiliev_06, Beardsley_Matsen_10} have shown that the dependence of $(\chi_e N)_{ODT}$ upon $L$ is modest for symmetric diblock copolymers in cells of this size or larger. 

\begin{figure}
\centering
\includegraphics[width=0.70\columnwidth]{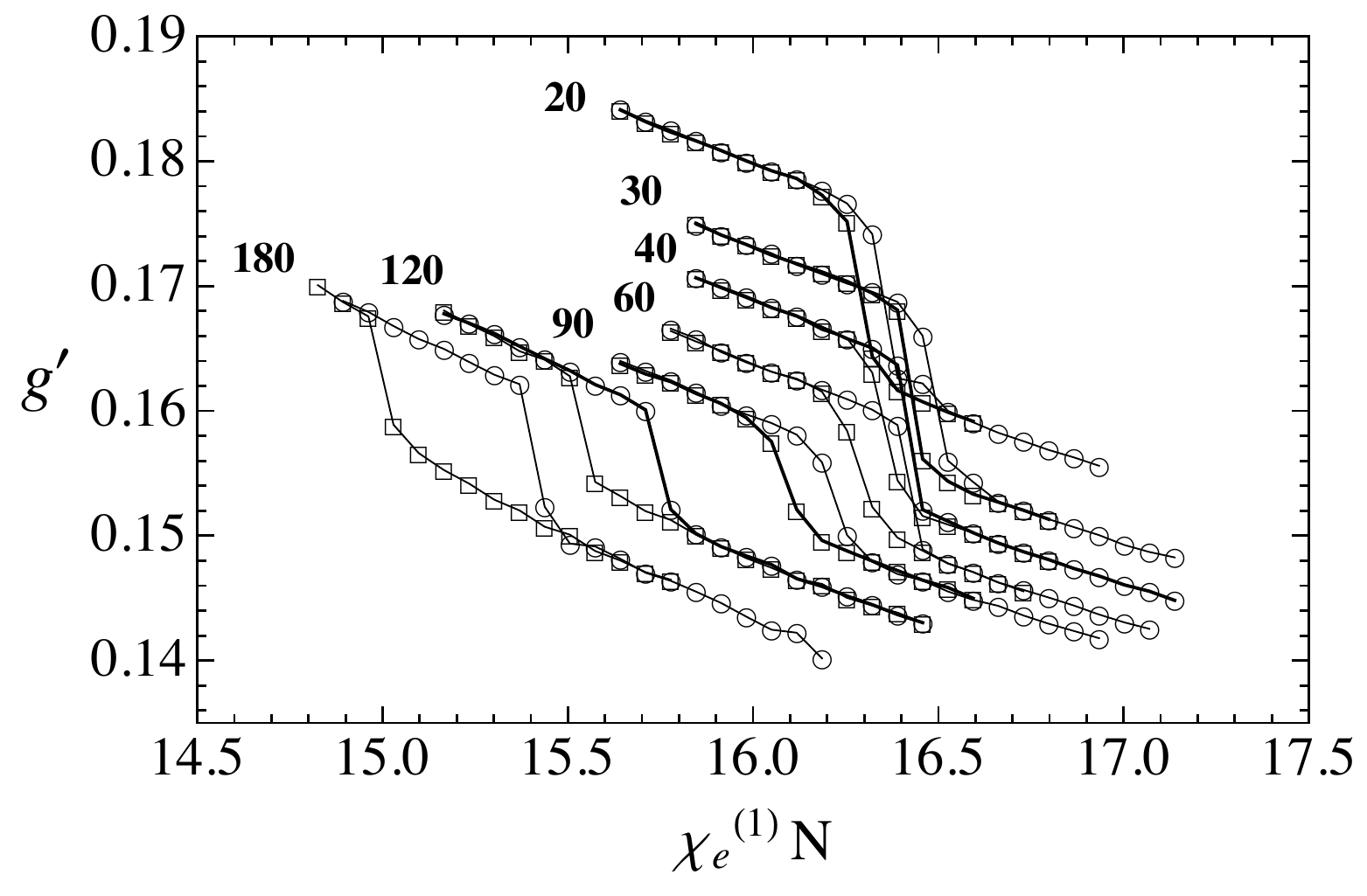} 
 \caption{Free energy derivative $g'$ \vs $\chi_e^{(1)} N$ for lattice model simulations with $N=20$ to $180$, where $\chi_{e}^{(1)}(\alpha) \equiv z_{\infty}\alpha/k_B T$ is a linear approximation for $\chi_{e}(\alpha)$. Values of $g'$ shown here were calculated using Eq. (2) of the main text, also using approximation $\chi_{e}^{(1)}$.  Results are show both for simulations that were initialized with all disordered configurations ($\bigcirc$) and simulations that were initialized with ordered configurations ($\square$). The persistence of metastable ordered and disordered phases very near the ODT is evident from the existence of metastability loops in which the results obtained from different initial states differ.}
 \label{fig:metastability-loops}
\end{figure}

In the simulations presented here, separate sets of parallel tempering simulations were conducted for each $N$, initialized from either (1) all disordered configurations or (2) all ordered configurations.  Results obtained from simulations with these different initial states were found to converge outside of a small range of values of $\alpha$ near the ODT. Near the ODT, however, results from different initial conditions differed, creating the metastability loops visible in Fig.~\ref{fig:metastability-loops}. The minimum and maximum values of $\alpha$ within these loops provide bounds on the true value $\alpha_{ODT}$ of $\alpha$ at the ODT. Our estimates of $\alpha_{ODT}$ for model F, as reported in Fig. 3 of the main text, are the midpoints of these metastability loops. The uncertainty $\Delta (\chi_e N)$ in the value of $\chi_e N$ at the ODT due to metastability is $\Delta (\chi_{e} N)\leq 0.1$ for $N=$ 20, \ldots, 120 and $\Delta (\chi_{e} N) \simeq 0.2$ for $N=180$. This uncertainty is too small to significantly affect the quality of the collapse shown in Fig. 3 of the main text.

\section{Bead-Spring Models}
Models H, S1, S2 and S3 are continuum bead-spring models, which are described in the main text. All bead-spring simulations reported here are molecular dynamics simulations that were conducted using the HOOMD-blue code \cite{Anderson2008} for GPU-accelerated MD simulations.  The simulations used to measure the structure factor $S(q)$ in the disordered phase, measure the average AB pair energy $\langle U_{AB} \rangle$ in the disordered phase, and locate the ODT were all isobaric, isothermal molecular dynamics simulations that were carried out using the integrator of Martyna, Tobias, and Klein \cite{Martyna_Tobias_Klein_94}, using a time step $\Delta t = 0.005$ in natural units ($k_{B}T=1$, $\sigma=1$, bead mass =1). These simulations were performed using a fixed set of values for potential energy parameters $\epsilon_{AA}=\epsilon_{BB}$, $r_{c}$, $\kappa$, and $l_{0}$ and the pressure $P$ for each model (non-dimensionalized in the same units), over a range of values of $\epsilon_{AB} = \epsilon_{AA} + \alpha$. Isothermal, constant volume (NVT) simulations of models H and S1 have been carried out previously \cite{Qin_Morse_12,Glaser_Morse_12,Glaser_Morse_14}, using the same potential energy parameters. Values for all simulation input parameters are given in Table 1.

The pressure used in NPT simulations of each such model was chosen so as to yield a predetermined target value $c$ for the monomer concentration in the limit $\alpha = 0$, $N \rightarrow \infty$ of infinite homopolymers. The target values of $c = 0.7 \sigma^{-3}$ for model H and $c = 3.0 \sigma^{-3}$ for model S are the concentrations used in earlier NVT simulations of these models \cite{Glaser_Morse_14}. The required pressure $P$ for each model was determined by running NVT simulations for each model for several values of $N$ in the $\alpha=0$ homopolymer state with a monomer concentration equal to the target value, and measuring the pressure. The pressure $P(N)$ was found to exhibit a nearly perfect linear dependence on $1/N$, $P(N) = P_{\infty} + \delta/N$. The extrapolated value $P = P_{\infty}$ was used for all NPT simulations.

NPT simulations of the bead-spring models were carried out using chains of length $N=$16, 32, 64, and 128. Several types of NPT simulation were performed for each such model and chain length. 

(1) Simulations of the disordered phase were performed in NPT ensemble using an $L \times L \times L$ cubic simulation cell. The number of molecules $M$ for each of these simulations was chosen, as in Ref.~\cite{Glaser_Morse_14}, to yield a length $L$ of approximately 3 times the RPA prediction $2\pi/q_{0}^{*}$ for the layer spacing evaluated at the ODT for a system with the asymptotic monomer concentration $c$.

(2) Simulations of the ordered phase were performed over a range of parameters near and above the ODT value of $\alpha$ using a tetragonal $L_x \times L_x \times L_z$ simulation cell in which $L_x$ and $L_z$ can fluctuate independently so as to allow the layer spacing to adjust to create a state of isotropic pressure. These simulations were initialized with artificially ordered configurations of 3 lamellar periods with layers oriented normal to the $z$ axis. In systems that did not disorder, the layers remained in this orientation. All reported thermodynamic and structural properties of the ordered phase, including $\langle U_{AB} \rangle$, the layer spacing $d$, and the composition profile $\phi_{A}(z)$, were obtained from such tetragonal NPT simulations.

Initial estimates of the value of $\alpha$ at the ODT were obtained from observations of spontaneous disordering of tetragonal NPT simulations of systems that were initialized with an ordered configuration, and of spontaneous ordering of cubic NPT simulations of systems that were initialized with a disordered configuration, for simulations with values of $\alpha$ spaced closely around the ODT.  Measurements of $\langle U_{AB} \rangle$ yielded metastability loops similar to those shown in Fig. 1 for the lattice model. Ordered and disordered phases were also identified by tracking the absolute magnitude of the Fourier amplitude $\psi(\vec q)$ of the order parameter $\psi(\vec r) = c_{A}(\vec r) - c_{B}(\vec r)$ at the wavevector $\vec q$ for which $|\psi(\vec q)|$ is maximum, corresponding to a Bragg peak in the ordered phase. These simulations yielded initial estimates of the values $\alpha_{\mathrm{ODT}}$ and $d_{\mathrm{ODT}}$ of $\alpha$ and the layer spacing $d$ at the ODT.

(3) Well-tempered metadynamics simulations were run for each model and chain length to more precisely locate the ODT by measuring the difference in free energy between the competing phases. These simulations are discussed in a separate section below.

\begin{table}
\begin{tabular}{|l|r|r|r|r|r|r|}
\hline\hline
model  & $\epsilon_{AA}$ & $r_{c}$ & $\kappa$ & $l_{0}$ & $c$ & P \\

\hline
H  & 1.0  & 1.1225 & 400.0 & 1.0 & 0.7 & 2.307 \\
S1 & 25.0 & 1.0    & 3.406 & 0.0 & 3.0 & 20.249 \\
S2 & 25.0 & 1.0    & 1.135 & 0.0 & 1.5 & 4.111 \\
S3 & 25.0 & 1.0    & 0.867 & 0.0 &1.5 & 4.132 \\
\hline\hline
\end{tabular} 
\caption{Simulation input parameters for the bead-spring models in units of $k_{B}T = 1$ and $\sigma = 1$. Here, $c$ denotes the extrapolated monomer concentration for infinite homopolymers at the specified pressure. \label{tab:input_params}}
\end{table}

\begin{table}
\begin{tabular}{|l|r|r|r|r|r|}
\hline\hline
model & M & L & $\alpha_{\mathrm{ODT}}$ & $(\chi_e N)_{\mathrm{ODT}}$ & $\Nbar$ \\
\hline
H-64  & 2026  & 57.00  & 1.587  & 22.86  & 240.2  \\
S1-16 & 2007  & 22.04  & 4.920  & 22.76  & 238.9  \\
S1-32 & 2555  & 30.09  & 2.220  & 19.57  & 477.7  \\
S1-64 & 3269  & 41.16  & 1.031  & 17.24  & 955.4  \\
S1-128 & 4162 & 56.21  & 0.478  & 15.33  & 1910.9 \\
S2-16 & 3137  & 32.22  & 14.678 & 17.44  & 955.5  \\
S2-32 & 4145  & 44.55  & 5.771  & 15.61  & 1911   \\
S2-64 & 5573  & 61.95  & 2.534  & 14.35  & 3822   \\
S3-16 & 3967  & 34.85  & 12.430 & 15.70  & 1907   \\
S3-32 & 5391  & 48.63  & 4.990  & 14.39  & 3815   \\
S3-64 & 7227  & 67.56  & 2.213  & 13.36  & 7629   \\
F-20  & 2500  & 50     & 0.167  & 27.48  & 90.13  \\
F-30  & 2880  & 60     & 0.112  & 25.80  & 135.2  \\
F-40  & 3144  & 68     & 0.084  & 24.30  & 180.3  \\
F-60  & 3413  & 80     & 0.056  & 22.44  & 270.4  \\
F-90  & 3691  & 94     & 0.037  & 20.59  & 405.6  \\
F-120 & 4199  & 108    & 0.027  & 18.98  & 540.8  \\
F-180 & 5111  & 132    & 0.017  & 17.44  & 811.2  \\
\hline\hline
\end{tabular}
\caption{Chain length $N$, number of molecules $M$, simulation cell size $L$, and precise estimates of ODT for all systems studied in this work. Cell size $L$ is given in terms of length unit of $\sigma$ = 1 for bead-spring models and $d$ = 1 for model F. Values of $M$ and $L$ given here for the bead-spring models are those used in well-tempered metadynamics simulations conducted to locate the ODT.}
\label{tab:systems}
\end{table}
  
\section{Parameter Calibration}

Here, we present the analysis used to estimate the model-dependent phenomenological parameters $b$, $z_{\infty}$ and $\chi_{e}(\alpha)$ that are required in our theoretical analysis. Fitting procedures for these quantities are identical to those presented in Ref.~\cite{Glaser_Morse_14}. Values for the resulting coefficients are given in Table \ref{tab:params}.

\subsection{Statistical Segment Length}
\begin{figure}
%\centering\includegraphics[width=0.98\columnwidth]{smFig2}
\centering\includegraphics[width=0.70\columnwidth]{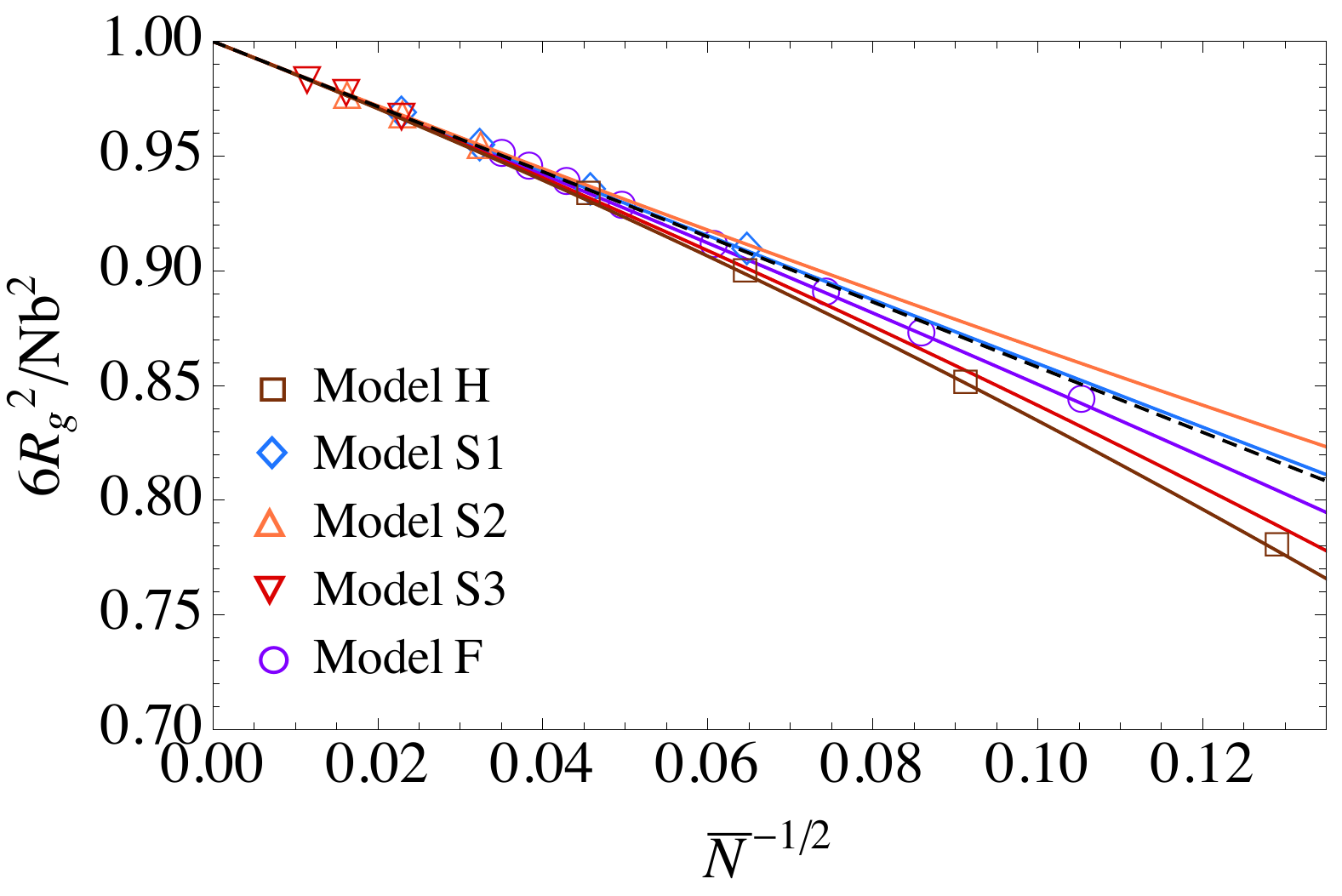}
\caption{Results for $6 R_g^2/Nb^2$ (symbols) vs. $\Nbar^{-1/2}$ in homopolymer melts, for all five models. Solid lines are fits to Eq. (\ref{eq:Rg_fit}), which were used to determine values for $b$. The dashed line is the predicted universal asymptote $1-1.42\Nbar^{-1/2}$, without the $1/N$ correction. \label{fig:b}}
\end{figure}

Values of the statistical segment length $b$ for each model were obtained, as in Ref.~\cite{Glaser_Morse_14}, by evaluating the radius of gyration $R_{g}(N)$ for homopolymers with several chain lengths and extrapolating to obtain the limit $b^{2} = \lim_{N \rightarrow \infty} 6 R_{g}^{2}(N)/N$.  Because this defines $b$ for continuum models as a property of a hypothetical system of infinite homopolymers with a specified concentration $c$ and a corresponding pressure $P$, it can be evaluated by extrapolating results of either NVT or NPT simulations. Values of $b$ for models S2, S3 and F were obtained by analyzing NVT homopolymer simulations with the target monomer concentrations $c = MN/V$ for several values of $N$, as was done for models H and S1 in Ref.~\cite{Glaser_Morse_14}.  The ROL theory predicts \cite{Qin_Morse_09, Glaser_Morse_14} that $6 R_{g}^{2}(N)/N$ should vary with $N$ in a homopolymer melt as
\begin{equation}
    \frac{6 R_{g}^{2}(N)}{N} \simeq 
    b^{2} \left [ 1 - \frac{1.42}{\Nbar^{1/2}} + \frac{\gamma}{N} \right ]
    \quad. \label{eq:Rg_fit}
\end{equation}
where the coefficient 1.42 is a theoretically predicted universal coefficient. Results for each model were fit to Eq. (\ref{eq:Rg_fit}) by treating $b$ and $\gamma$ as fitting parameters. Fig.~\ref{fig:b} shows the resulting fits for all five models. 

\subsection{Effective Coordination Number}
\begin{figure}
%\centering\includegraphics[width=0.98\columnwidth]{smFig3}
\centering\includegraphics[width=0.70\columnwidth]{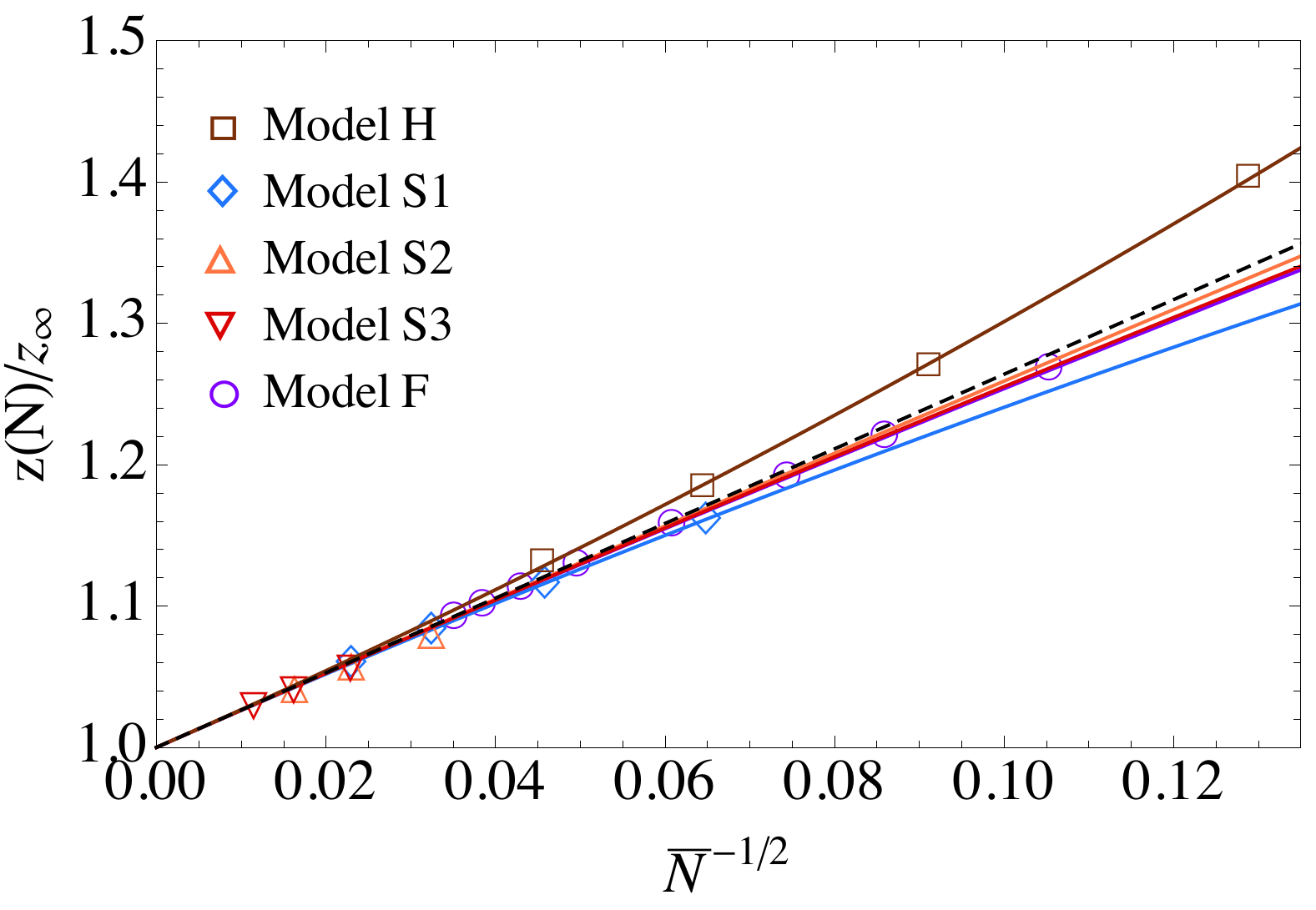}
\caption{Results for $z(N)/z_\infty$ (symbols) vs. $\Nbar^{-1/2}$ for all five models. Solid lines are fits to Eq. (\ref{eq:z_fit}). The dashed line is the predicted asymptote $ 1 + (6/\pi)^{3/2}\bar{N}^{-1/2}$. \label{fig:z}}
\end{figure}

It was shown in Ref.~\cite{Morse_Chung_09} that the RPA interaction parameter $\chi_{e}$ for structurally symmetric models such as those studied here is given to first order in $\alpha$ by $\chi_{e}(\alpha) \simeq z_{\infty}\alpha/k_{B}T$, where the coefficient $z_{\alpha}$ is an ``effective coordination number" that can be extracted from homopolymer simulations. This coefficient is given by the $N \rightarrow \infty$ limit of a quantity $z(N) = \langle U_{inter} \rangle / (MN\epsilon)$, where $U_{inter}$ is the inter-molecular pair interaction energy in a homopolymer melt of $M$ chains of length $N$ with a pair potential $V(r) = \epsilon u(r)$ for bead-spring models, or a nearest-neighbor interaction of strength $\epsilon$ in a lattice model.  It was also predicted \cite{Morse_Chung_09} that $z(N)$ should vary with $N$ as
\begin{equation}
    z(N) \simeq z_{\infty}
    \left [ 1 + \frac{(6/\pi)^{3/2}}{\Nbar^{1/2}} + \frac{\delta}{N} \right ]
    \quad, \label{eq:z_fit}
\end{equation}
where $z_{\infty}$ and $\delta$ are model-dependent parameters. Values of $z_{\infty}$ were determined for models S2, S3 and F, as done previously for models H and S1 \cite{Glaser_Morse_14}, by running NVT homopolymer simulations at the target concentration $c$ for several chain lengths, and fitting the results for each model to Eq. (\ref{eq:z_fit}), using $z_{\infty}$ and $\delta$ as fitting parameters. Fig.~\ref{fig:z} shows the resulting fits. Values are given table \ref{tab:params}.

\subsection{Estimating $\chi_e(\alpha)$ by fitting $S(q)$}
A nonlinear approximation for $\chi_e(\alpha)$ was estimated for each model by fitting simulation results for the structure factor $S(q)$ in the disordered phase to predictions of the ROL theory \cite{Grzywacz_Morse_07, Qin_Morse_11}, as done previously in Ref.~\cite{Glaser_Morse_14}. Bead-spring simulations used for this purpose were cubic NPT simulations carried out at the same pressure as that used in all other simulations. Because our previous simulations of models H and S1 were NVT simulations, we have obtained new data and new fits for these models using NPT simulations. In order to obtain reliable data for $S(q)$ for each model over a range of values of $\alpha$ large enough to reach the ODT for the shortest chains of interest (\ie for $N=16$ for models S1, S2, and S3 or $N=20$ for model F),  we included data for models S1, S2, S3 and F from additional simulations of chains of length $N=12$ in the data set that was used to estimate $\chi_e(\alpha)$. These additional simulations of short chains were not used for other purposes, and are not reported in the main text of this article.

\begin{figure}

%Twocolumn
%\centering\includegraphics[width=0.70\columnwidth]{smFig4a}\\
%\centering\includegraphics[width=0.70\columnwidth]{smFig4b}\\
%\centering\includegraphics[width=0.70\columnwidth]{smFig4c}\\
%\centering\includegraphics[width=0.70\columnwidth]{smFig4d}

%Preprint
\centering\includegraphics[width=0.5\columnwidth]{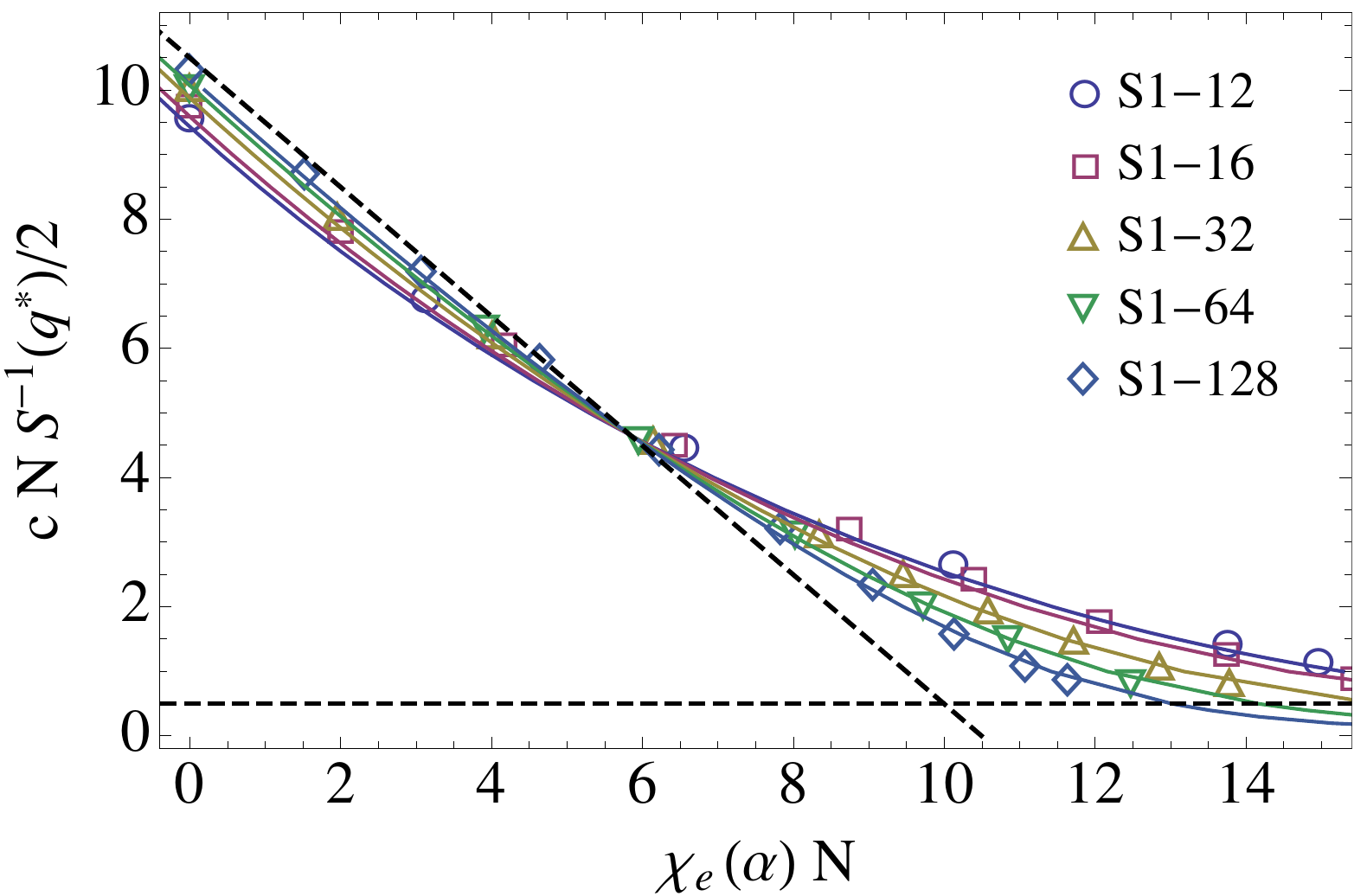}\\
\centering\includegraphics[width=0.5\columnwidth]{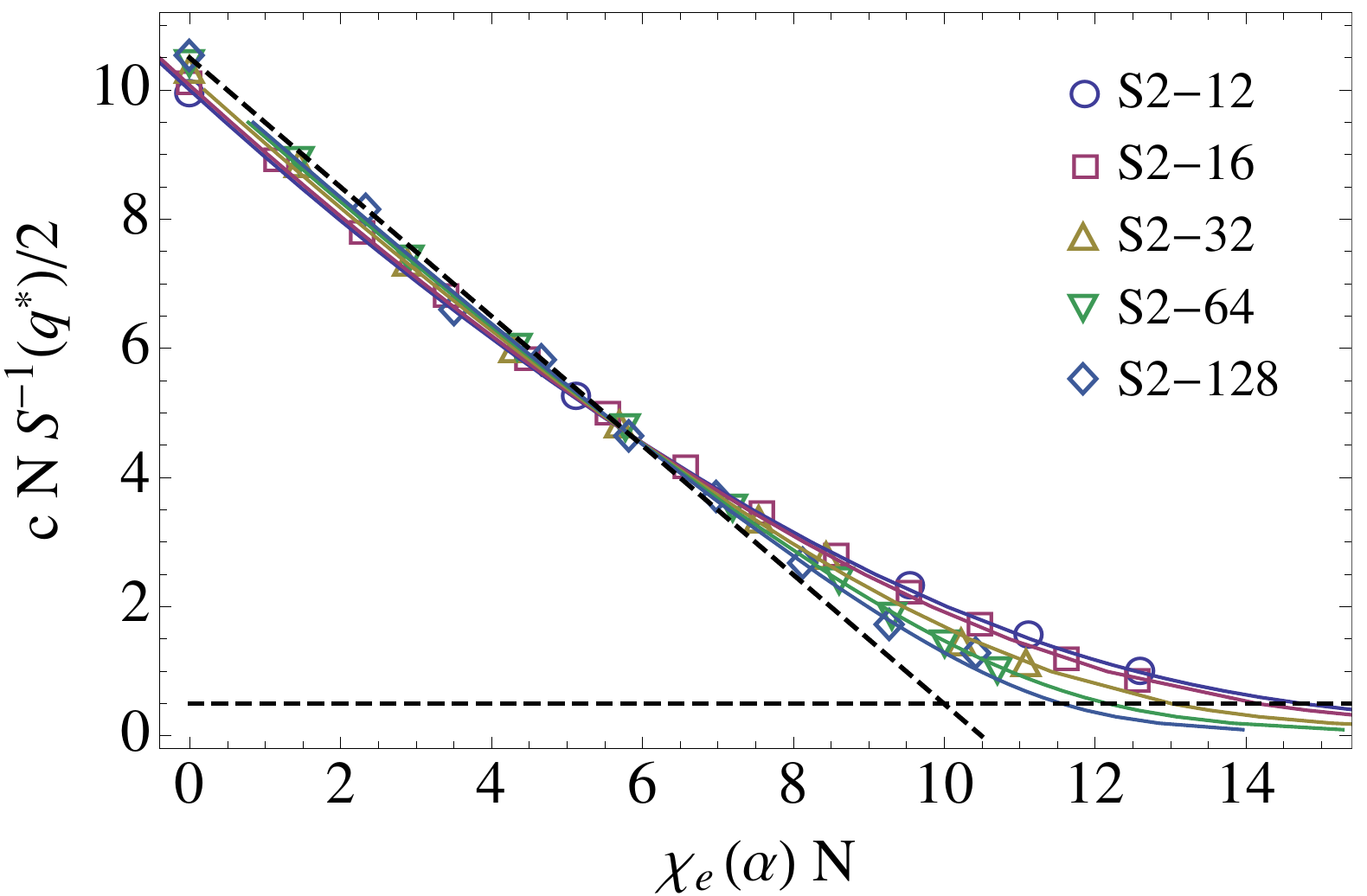}\\
\centering\includegraphics[width=0.5\columnwidth]{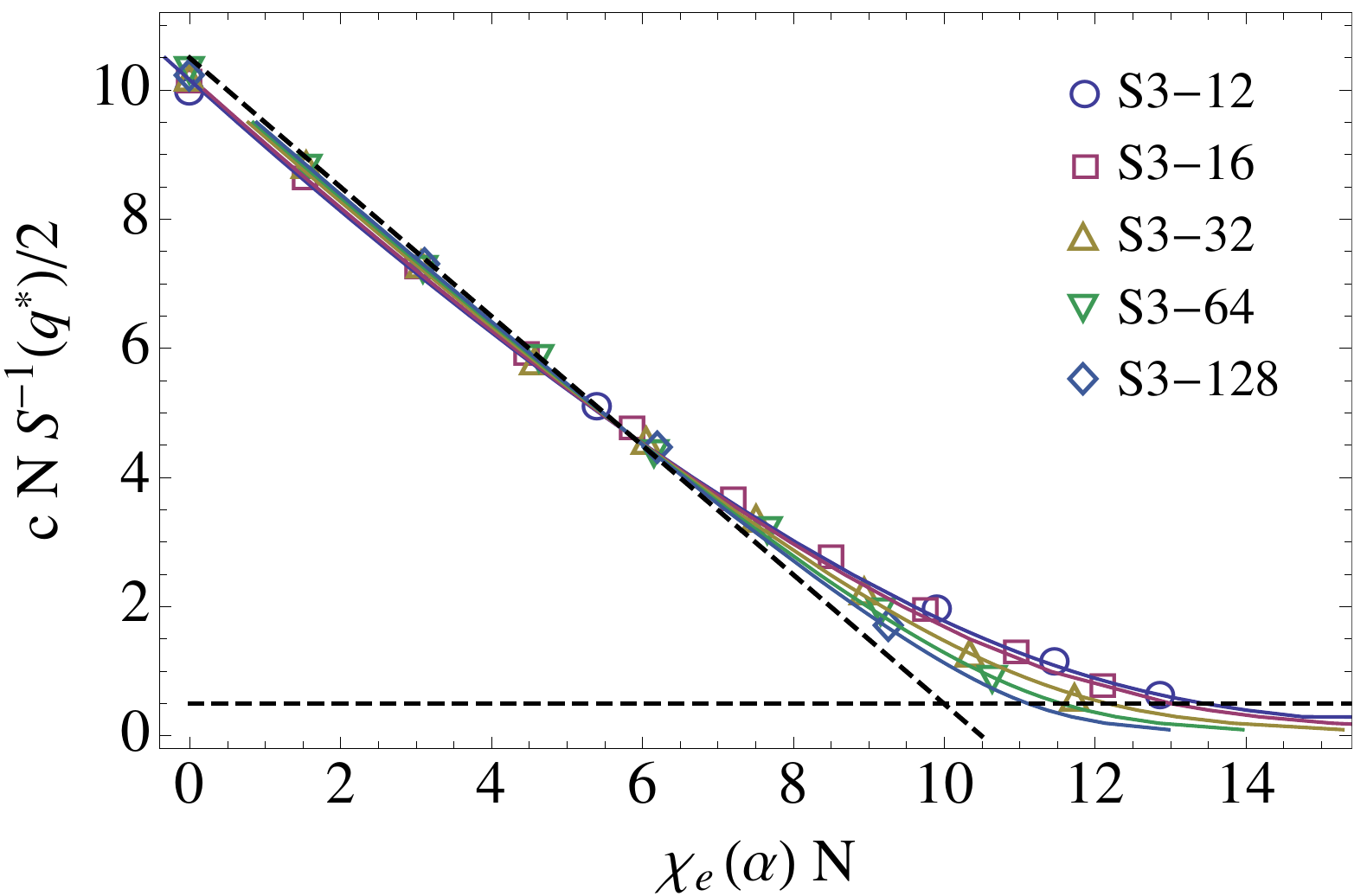}\\
\centering\includegraphics[width=0.5\columnwidth]{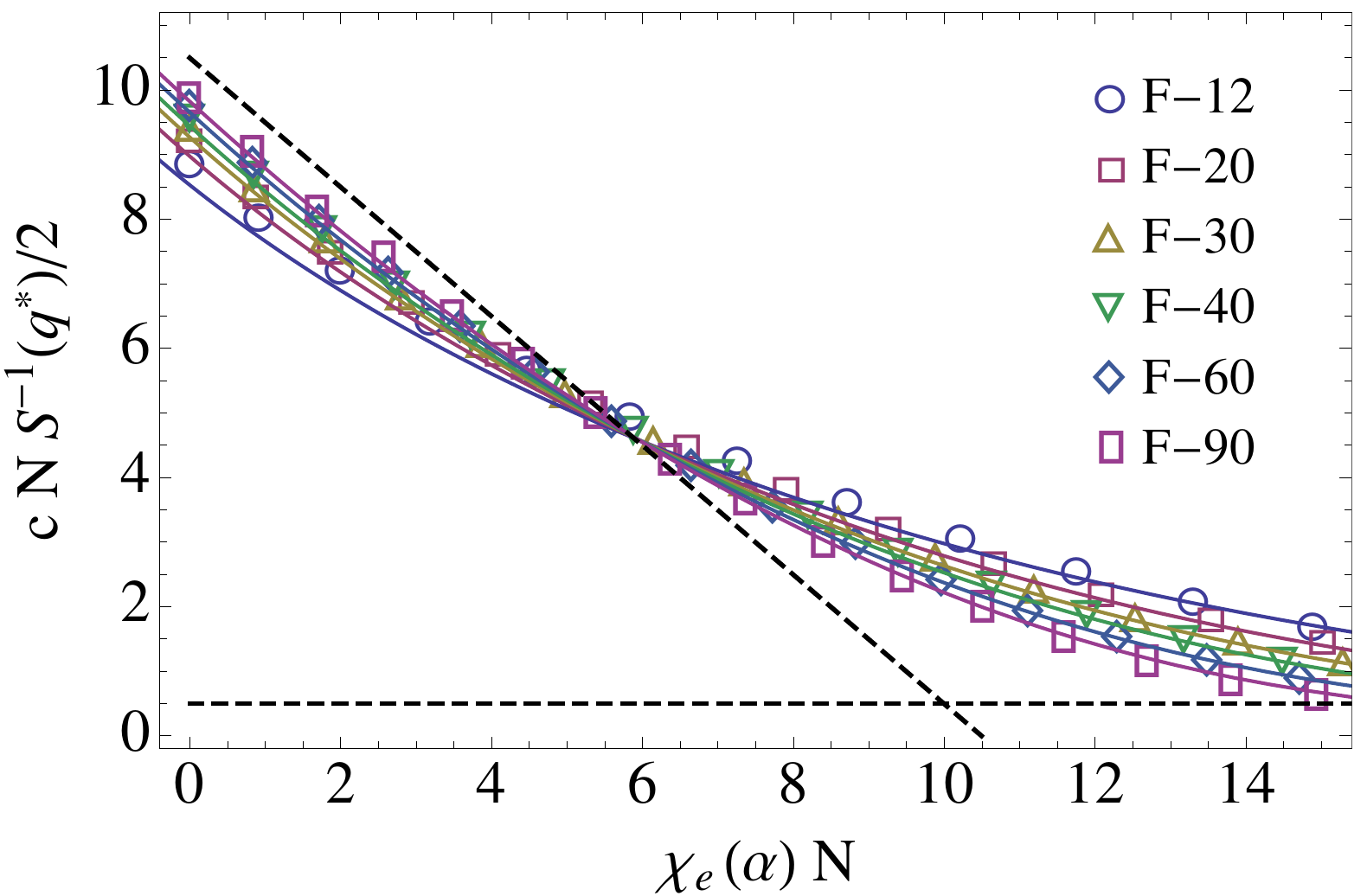}

\caption{Fits of the inverse peak structure factor $cNS^{-1}\left(q^*\right)$ in the disordered phase to the one-loop theory \cite{Grzywacz_Morse_07, Qin_Morse_11} to estimate $\chi_e(\alpha)$ for models S1, S2, S3, and F \label{fig:Sinv_fit}}
\end{figure}

\begin{figure}
%\centering\includegraphics[width=0.98\columnwidth]{smFig5}
\centering\includegraphics[width=0.70\columnwidth]{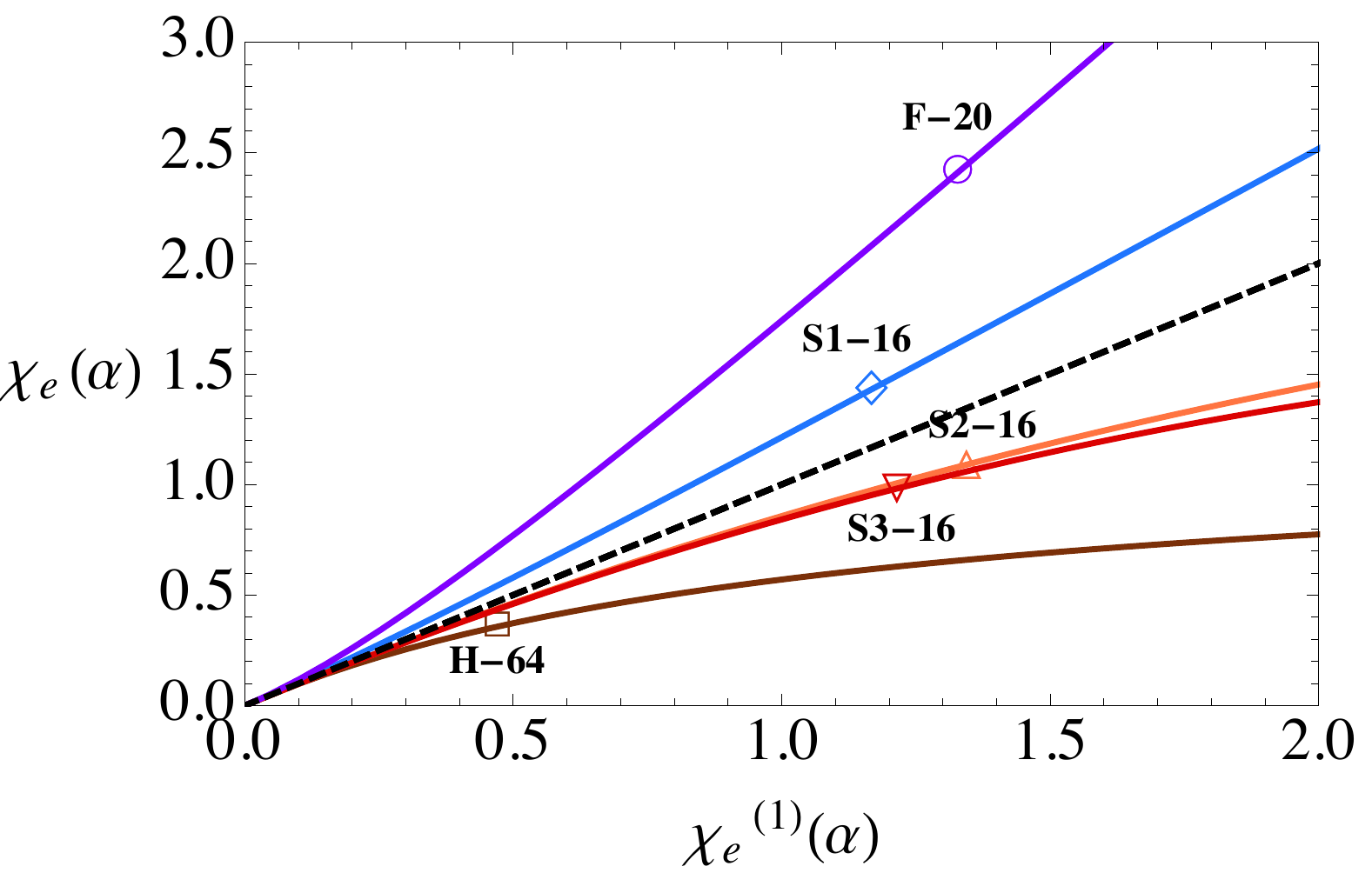}
\caption{ $\chi_{e} (\alpha)$ obtained from fits to the one-loop theory plotted vs. linear approximation $\chi_{e}^{(1)} \equiv z_{\infty} \alpha / k_B T$, for all five models. The symbols indicate values of $\chi_e$ at the ODT for the shortest chain length for which we have located the ODT for each model. The dashed line shows the linear approximation $\chi_e(\alpha) = \chi_{e}^{(1)}$. \label{fig:chie} }
\end{figure}

An approximation for $\chi_{e}(\alpha)$ for each model was obtained by a simultaneous fit of ROL theory predictions to simulation results for $cNS^{-1}(q^{*})$ from several different chain lengths, using the same function for $\chi_{e}(\alpha)$ for all chain lengths. The fit assumed the functional form
\begin{equation}
    \chi_{e}(\alpha) = \frac{z_{\infty} \hat{\alpha} + a_2 \hat{\alpha}^{2}}
                            {1 +  d_1 \hat{\alpha} + d_2 \hat{\alpha}^{2}}
     \label{eq:chie_fit}
\end{equation}
for each model, where $\hat{\alpha} = \alpha / k_{B}T$, with $d_2 = 0$ in all models except model H. The coefficients $a_2$, $d_1$ and (for model H) $d_2$ are treated as fitting parameters. The use of $z_{\infty}$ as the leading term in the numerator constrains the fit to agree with the results of perturbation theory \cite{Morse_Chung_09}. The functional form for model H, for which $d_2 \neq 0$, was chosen to make $\chi_{e}(\alpha)$ approach a finite limit as $\alpha \rightarrow \infty$, for physical reasons that are discussed in Ref.~\cite{Glaser_Morse_14}. The form used for all other models (with $d_2 = 0$) yields a linear increase for large $\alpha$. Results of the fits for models S1, S2, S3 and F are shown in Fig, \ref{fig:Sinv_fit}. Model H is not shown because it is very similar to the fit shown for NVT simulations in Ref.~\cite{Glaser_Morse_14}, and because we evaluated the ODT for this model only for chains of length $N=64$, for which the linear approximation almost suffices. Resulting values for the coefficients $a_2$, $d_1$ and (for model H) $d_2$ are given in Table \ref{tab:params}. Plots of the resulting estimates of $\chi_e(\alpha)$ are shown in Fig.~\ref{fig:chie}.

\begin{table}
\begin{tabular}{|l|r|r|r|r|r|r|r|r|r|}
\hline\hline
model & $b$ & $(cb^3)^2$ & $z_\infty$ & $a_2$ & $d_1$ & $d_2$ \\
\hline
H  & 1.404 & 3.753  & 0.2965 & 2.56     & 8.20    & 2.123 \\
S1 & 1.088 & 14.93  & 0.237  & 0.138    & 0.438   & 0 \\
S2 & 1.727 & 59.70  & 0.0916 & -0.00087 & 0.00420 & 0 \\
S3 & 1.938 & 119.21 & 0.0977 & -0.00144 & 0.00086 & 0 \\
F  & 1.745 & 4.506  & 4.897  & 88.5     & 8.30    & 0 \\
\hline\hline
\end{tabular}
\caption{Estimates of parameters $b$ and $z_{\infty}$ and the coefficients in Eq. (\ref{eq:chie_fit}) for $\chi_{e}(\alpha)$ for all five models, as determined by fitting simulation results. The statistical segment length $b$ is given in units with $\sigma = 1$ for bead-spring models or $d = 1$ for the FCC lattice model. The value of $(cb^{3})^2$ for NPT bead-spring simulations is calculated using the target value of $c$ for infinite homopolymers.\label{tab:params}}
\end{table}

\section{Well-Tempered Metadynamics}

The well-tempered metadynamics algorithm \cite{Laio2002, Barducci2008} (WTMetaD) was used to precisely locate the ODT of the bead-spring models (models H, S1, S2 and S3). Metadynamics (MetaD) is an adaptive bias potential technique that is based on the simulation of a modified Hamiltonian that is given by the sum
\begin{equation}
   H = H_{0} + V(\Psi)
\end{equation}
of the physical system Hamiltonian $H_{0}$ and a fictitious potential $V(\Psi)$ that depends on a collective variable $\Psi$. Like other adaptive biasing techniques, such as Wang-Landau sampling, WTMetaD causes $V(\Psi)$ to evolve during the simulation in a manner that is designed to lower free energy barriers between different macroscopic states. WTMetaD provides an estimate of the free energy $G(\Psi)$ as a function of the collective variable, in which (for an appropriate choice of collective variable) different macroscopic states show up as distinct minima in $G(\Psi)$. 

The effectiveness of metadynamics depends critically upon the choice of an appropriate collective variable. Natural choices of a collective variables for a crystallization transition can be expressed in terms of the Fourier amplitudes of the order parameter $\psi(\vec r) = c_{A}(\vec r) - c_{B}(\vec r)$. These can be expressed as sums
\begin{equation}
  \psi(\vec q) \equiv  \frac{1}{NM}
  \sum_{j=1}^{MN} \epsilon_{j} e^{i {\vec q}\cdot{\vec r}_{j}}
\end{equation}
where ${\vec q}$ is a wavevector commensurate with the simulation cell, $\sum_{j}$ is a sum over all $NM$ monomers in a system of $M$ chains of length $N$, ${\vec r}_{j}$ is the position of monomer $j$, and $\epsilon_{j} = \pm 1$ is a prefactor of $+1$ for A monomers and $-1$ for B monomers. We experimented with collective variables that depended only upon the Fourier amplitudes for the Bragg peaks of the lamellar phase in some expected orientation, but encountered difficulties due to the tendency of the system to order in unexpected orientations. In response, we adopted a multi-mode collective variable defined as a regularized sum
\begin{equation}
\Psi \equiv \left [ \sum_{\vec{q}} |\psi(\vec q)|^{n} f (q / q_{\mathrm{cut}} ) \right ]^{1/n}
\label{eq:order_parameter}
\end{equation}
over all accessible wavevectors $\vec{q}$, and experimented with different possible values of the integer $n$. Here, $f(q/q_{\mathrm{cut}})$ is a cutoff function that is introduced to suppress contributions from $q \gg q^{*}$, and the sum is over all allowed wavevectors for which $f(q/q_{\mathrm{cut}})$ is not negligible. We took $f(x)$ to be a modified Fermi function $f(x) = \left\{1+ \exp\left[12\left(x - 1\right)\right]\right\}^{-1}$, with $q_{\mathrm{cut}}$ slightly larger than the peak wavenumber $q^{*}$. 

\begin{figure}
\includegraphics[width=0.70\columnwidth]{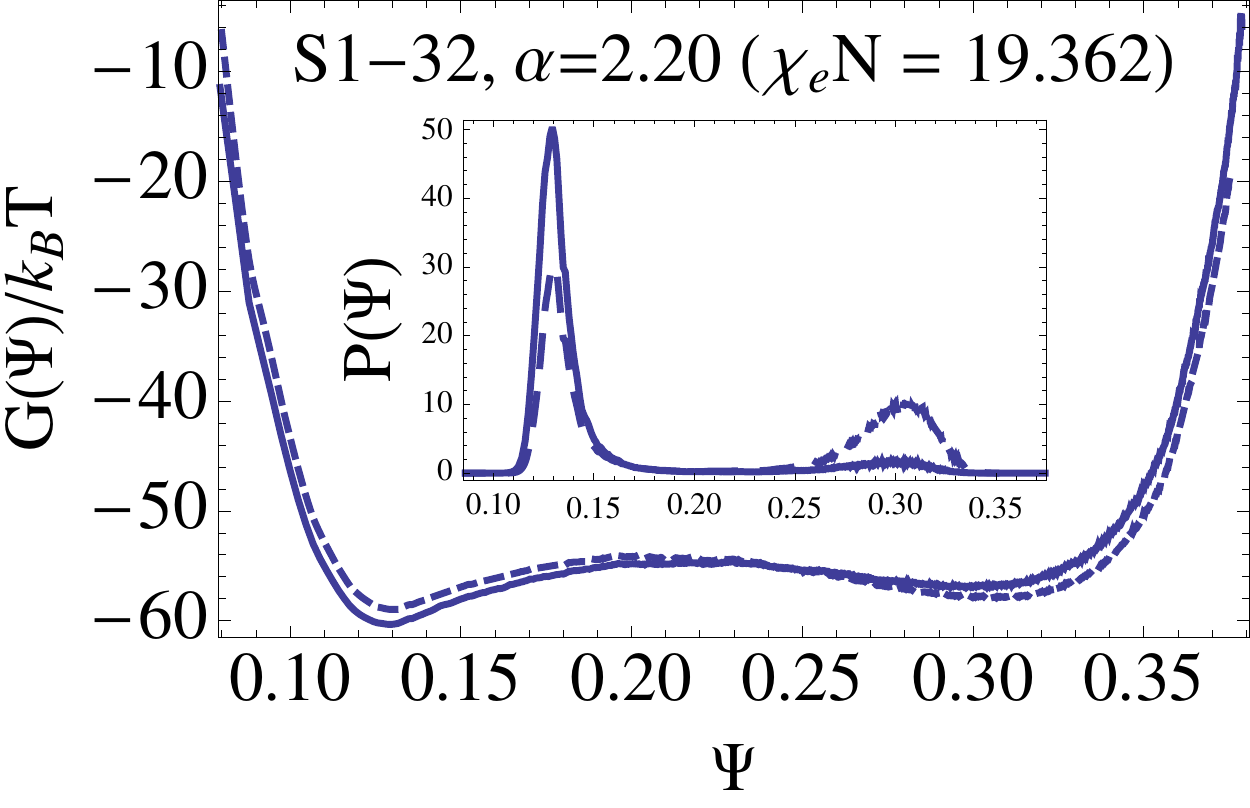}
\caption{Constrained Gibbs free energy $G(\Psi)$ from WTMetaD simulations as function of the order parameter $\Psi$ for model
S1-32 (M=2543, 3 layers) calculated at $\alpha=2.20$ (solid curve), and extrapolated to the ODT at $\alpha_{\mathrm{ODT}}=2.213$ (dashed curve) using linear extrapolation, Eq.~(\ref{eq:extrap}). {\em Inset}: Equilibrium probability distribution $P(\Psi) \propto e^{-G(\Psi)/k_B T}$ for the same two values of $\alpha$ (solid/dashed curves). The ODT (dashed curve) was determined by an equal area rule. \label{fig:G_WTMetaD}}
\end{figure}

Eq. (\ref{eq:order_parameter}) defines a regularized norm for composition fluctuations. The choice $n=2$, corresponding to a Euclidean norm, does not adequately discriminate the ordered and disordered phases. Increasing $n$ increases the relative weight of the wavevector $\vec{q}$ for which $|\psi(\vec q)|$ is maximum, and thereby increases the separation between the values of $\Psi$ in the ordered and disordered phases.  The choice $n=4$ was found to yield adequate discrimination, and was used in all calculations. An example of the resulting converged free energy $G(\Psi)$ is shown in Fig.~\ref{fig:G_WTMetaD}.

%Fig.~\ref{fig:validation_twolayers} shows a validation of our metadynamics algorithm in which we compare the constrained free energy $G(\psi)$ obtained at a value of $\alpha$ very near the ODT to the free energy inferred from the probability distribution in a completely unbiased MD simulation of the same system. The disordered and ordered phases correspond to the minimum at small and large values of $\Psi$, respectively. The system used for this test (model S1-16 in a system large enough to accommodate only two layers) was chosen because it was found to yield a free energy barrier small enough to allow ergodic sampling of both phases near the ODT in a conventional unbiased MD simulation, thus allowing a comparison of results from biased and unbiased simulations.  Calculated free energy barriers are found depend strongly on system volume, and do not allow ergodic sampling of unbiased simulations of both phases at the ODT of systems with three or more layers.  

To reduce the number of WTMetaD simulations necessary to identify $\alpha_{\mathrm{ODT}}$, we implemented an extrapolation scheme that allows one to estimate changes in the constrained free energy $G(\Psi, \alpha)$ with small changes of $\alpha$. Given the results of a metadynamics simulation at a single value $\alpha = \alpha_{0}$, one can estimate $G(\Psi, \alpha)$ over a range of nearby values using a linear extrapolation
\begin{equation}
  G(\Psi, \alpha) \simeq 
  G(\Psi, \alpha_0) + \frac{\partial G}{\partial \alpha}\Big|_{\Psi, \alpha = \alpha_0}(\alpha - \alpha_0) 
  % + \mathcal{O}(\Delta \alpha^2),
\label{eq:extrap}
\end{equation}
in which the partial derivative is given by an average
\begin{equation}
   \frac{\partial G}{\partial \alpha}\Big|_{\Psi, \alpha = \alpha_0} =
   \frac{\langle U_{AB} \rangle\big|_{\Psi, \alpha_0}}{\epsilon_{AB}(\alpha_{0})}
\end{equation}
that can be evaluated for all values $\Psi$ over the course of a biased simulation. This scheme made it possible for us to identify the ODT for most systems from the results of a single WTMetaD simulation at a well chosen value of $\alpha$ near the true ODT. Fig.~\ref{fig:G_WTMetaD} shows an example of an extrapolation that was used to refine an accurate initial guess for $\alpha_{\mathrm{ODT}}$ for model S1-32. 

Metadynamics simulations were carried out in NPT ensemble using a cubic $L \times L \times L$ unit cell in which the number of molecules $M$ was chosen so as yield a commensurate cell for a 3 layer system oriented in a \{300\} orientation, using our best estimate of the equilibrium layer spacing and monomer concentrations in the ordered phase at the ODT from tetragonal NPT simulations conducted very near the ODT. With this choice of system size, ordered configurations observed in biased simulations almost all oriented along \{300\} or \{221\} orientations, which yield the same layer spacing $d = L/3$ in a cubic box. The value of $\alpha_{\mathrm{ODT}}$ for each system was identified using an equal area construction, requiring that regions near the ordered and disordered minima in $G(\Psi)$ yield equal contributions to the integral $\int d\Psi \; e^{-G(\Psi, \alpha)/kT}$.

Because our choice of collective variable is expressed as a sum of Fourier modes, efficient implementation of this metadynamics algorithm for large systems required the implementation of a particle-mesh algorithm similar to that used to treat Coulomb interactions in MD simulations \cite{Hockney1988}.  The WTMetaD algorithm was implemented as a publically accessible Integrator plug-in \cite{metad-plugin} to the HOOMD-blue simulation framework \cite{Anderson2008}. The particle-mesh scheme used to compute the order parameter and the forces derived from it have been fully implemented on the GPU using the CUFFT library, as part of the same plug-in. The implementation also runs on multiple GPUs, using a distributed FFT algorithm \cite{Bisseling2004,dfftlib}, which was needed for some larger systems. Further details of the implementation and testing of this algorithm will be discussed in a separate publication \cite{Glaser_Morse_14b}.

%\bibliography{bib/library,bib/polySim,bib/polyTheory,bib/supp,bib/smNote}
